\documentclass[12pt]{umj}
\usepackage[cp1251]{inputenc}
\usepackage[english]{babel}
\usepackage{amsmath}
\usepackage{amssymb}
\usepackage{amsfonts}
\usepackage{graphicx}
\usepackage{cite}
\usepackage[unicode]{hyperref}
\usepackage{cmap}
\def\udcs{517.929}

\subjclass{\udcs}

\def\negskp{\unskip\kern -2pt\unskip}

\newtheorem{theorem}{Theorem}

\newtheorem{lemma}{Lemma}

\def\eq#1{\begin{equation}#1\end{equation}}

\def\eqs#1{\begin{equation}\begin{split}#1\end{split}\end{equation}}

\def\seqs#1{\begin{equation*}\begin{split}#1\end{split}\end{equation*}}

\def\seq#1{\begin{equation*}#1\end{equation*}}
\def\qed{\vrule height0.6em width0.3em depth0pt} 

\def\ord{\hbox{ord}\,}

\def\eq#1{\begin{equation}#1\end{equation}}
\def\eqs#1{\begin{equation}\begin{split}#1\end{split}\end{equation}}
\def\seqs#1{\begin{equation*}\begin{split}#1\end{split}\end{equation*}}
\def\seq#1{\begin{equation*}#1\end{equation*}}
\def\qed{\vrule height0.6em width0.3em depth0pt}
\begin{document}
\thispagestyle{empty}
\title[Examples of Darboux Integrable Discrete Equations ...]
 {Examples of Darboux Integrable Discrete Equations Possessing First Integrals of an Arbitrarily High Minimal Order}

\author{R.N. Garifullin }
\address{Rustem Nailevich Garifullin, 
\newline\hphantom{iii} Ufa Institute of Mathematics, Russian Academy of Sciences,
\newline\hphantom{iii} 112 Chernyshevsky Street, 
\newline\hphantom{iii} Ufa 450008, Russian Federation
\newline\hphantom{iii}\url{http://matem.anrb.ru/garifullinrn}}
\email{rustem@matem.anrb.ru}

\author{R.I. Yamilov }
\address{Ravil Islamovich Yamilov, 
\newline\hphantom{iii} Ufa Institute of Mathematics, Russian Academy of Sciences,
\newline\hphantom{iii} 112 Chernyshevsky Street, 
\newline\hphantom{iii} Ufa 450008, Russian Federation
\newline\hphantom{iii} \url{http://matem.anrb.ru/en/yamilovri}}
\email{RvlYamilov@matem.anrb.ru}

\thanks{\sc R.N. Garifullin and R.I. Yamilov
Examples of Darboux Integrable Discrete Equations Possessing First Integrals of an Arbitrarily High Minimal Order.}
\thanks{\copyright \  R.N. Garifullin and R.I. Yamilov 2012}
\maketitle
{
\small
\begin{quote}
\noindent{\bf Abstract. } We consider a discrete equation, defined on the two-dimensional square lattice, which is linearizable, namely, of the Burgers type and depends on a parameter $\alpha$. For any natural number $N$ we choose $\alpha$ so that the equation becomes Darboux integrable and the minimal orders of its first integrals in both directions are greater or equal than $N$. 
\medskip

\medskip 

\end{quote}
}

\section{Introduction}

The most general form of the discrete Burgers equation introduced in \cite{rj06} reads:
\eqs{(u_{n+1,m+1}-&\beta_{n+1,m})(\alpha_{n,m}u_{n,m}+\gamma_{n,m})u_{n,m+1}=(u_{n,m+1}-\beta_{n,m})\\ *&(\alpha_{n+1,m}u_{n+1,m}+\gamma_{n+1,m})u_{n,m}, \ \ \alpha_{n,m},\beta_{n,m},\gamma_{n,m}\neq0.\label{ur_bur_o}}
Here $n,m$ are arbitrary integers, $\alpha_{n,m},\beta_{n,m},\gamma_{n,m}$ are known complex parameters, while $u_{n,m}$ is an unknown complex-valued function. Eq. (\ref{ur_bur_o}) can be obtained by the discrete Hopf-Cole transform (see e.g. \cite{l83})
\eq{u_{n,m}=\frac{v_{n+1,m}}{v_{n,m}}\label{hopf-cole}} from the following non-autonomous linear equation:
\eq{v_{n+1,m+1}=\alpha_{n,m}v_{n+1,m}+\beta_{n,m}v_{n,m+1}+\gamma_{n,m}v_{n,m}.\label{bur_lin}}
In the completely autonomous case, the discrete Burgers equation \eqref{ur_bur_o} can be rewritten in the form:
\eqs{(u_{n+1,m+1}-&\beta)(u_{n,m}+\gamma)u_{n,m+1}\\=&(u_{n,m+1}-\beta)(u_{n+1,m}+\gamma)u_{n,m}\label{ur_bur_aut}.} This equation was known \cite{h04} earlier than \eqref{ur_bur_o}. It has been noticed in \cite{gy12} that there is one more autonomous particular case of eq. \eqref{ur_bur_o}, namely, the equation \eqs{(u_{n+1,m+1}-&\beta)(u_{n,m}+\gamma)u_{n,m+1}\\=&\alpha(u_{n,m+1}-\beta)(u_{n+1,m}+\gamma)u_{n,m}\label{ur_bur1}} generalizing \eqref{ur_bur_aut}. Unlike eq. \eqref{ur_bur_aut}, the last equation \eqref{ur_bur1} is related by \eqref{hopf-cole} to a non-autonomous linear equation.

In this paper we consider the particular case $\beta=\gamma$ of eq. \eqref{ur_bur1} which, after a rescale, can be expressed in the form: 
\eqs{(u_{n+1,m+1}+1)(u_{n,m}-1)u_{n,m+1}=\alpha(u_{n,m+1}+1)(u_{n+1,m}-1)u_{n,m}\label{ur_bur}.} Here $\alpha\neq0$ is a complex parameter. Corresponding to eq. \eqref{ur_bur} particular case of the linear equation \eqref{bur_lin} is: \eqs{v_{n+1,m+1}+v_{n,m+1}=\alpha^n(v_{n+1,m}-v_{n,m})\label{lin_eq}.} Our aim is to show that there exist equations of the form \eqref{ur_bur} possessing first integrals of an arbitrarily high minimal order.

Discrete equations of the form \eq{u_{n+1,m+1}=f(u_{n,m},u_{n+1,m},u_{n,m+1}),\label{dis_eq}} defined on the two-dimensional square lattice, are analogues of the hyperbolic equations \eq{u_{xy}=F(x,y,u,u_x,u_y)\label{hyp}.} There is an example similar to eq. \eqref{ur_bur} in the class \eqref{hyp}, see \cite{ZS01} and a discussion at the very end of the present paper.

\section{Definitions}

An equation of the form \eqref{dis_eq} is Darboux integrable if it has two first integrals $W_1,W_2$:
\eq{\label{darb1}(T_1-1)W_2=0,\quad W_2=w_{n,m}^{(2)}(u_{n,m+l_2},u_{n,m+l_2+1},\ldots,u_{n,m+k_2}),}
\eq{\label{darb2}(T_2-1)W_1=0,\quad W_1=w_{n,m}^{(1)}(u_{n+l_1,m},u_{n+l_1+1,m},\ldots,u_{n+k_1,m}).}
Here $l_1,l_2,k_1,k_2$ are integers, such that $l_1<k_1, \ l_2<k_2$, and $T_1,T_2$ are operators of the shift in the first and second directions, respectively: $T_1 h_{n,m}=h_{n+1,m}$, $T_2 h_{n,m}=h_{n,m+1}$.
We suppose that the relations (\ref{darb1},\ref{darb2}) are satisfied identically on the solutions of the corresponding equation \eqref{dis_eq}. The form of $W_1, W_2$ given in (\ref{darb1},\ref{darb2}) is the most general possible form. The functions $u_{n+i,m+j},$ with $i,j\neq0,$ are expressed in terms of $u_{n+i,m},\ u_{n,m+j}$ by using eq. \eqref{dis_eq}. The dependence of $W_1$ on $u_{n,m+j},j\neq0,$ and of $W_2$ on $u_{n+i,m},i\neq0,$ is impossible. We will call $W_1$ the first integral in the first (or $n$) direction, and $W_2$ will be called the first integral in the second (or $m$) direction.

It is obvious that, for any first integrals $W_1,W_2$, arbitrary functions $\Omega_1,\Omega_2$ of the form \eqs{\label{arb_fun}\Omega_1=\Omega_1(W_1,T_1^{\pm1}W_1,\ldots,T_1^{\pm j_1}W_1),\\ \Omega_2=\Omega_2(W_2,T_2^{\pm1}W_2,\ldots,T_2^{\pm j_2}W_2)} are also the first integrals. In particular, using the shifts, we can rewrite the first integrals $W_1,W_2$ of (\ref{darb1}, \ref{darb2}) as:
\eqs{W_1=w_{n,m}^{(1)}(u_{n,m},u_{n+1,m},\ldots,u_{n+k,m}),\ \ k\geq1,\\ W_2=w_{n,m}^{(2)}(u_{n,m},u_{n,m+1},\ldots,u_{n,m+l}),\ \ l\geq1.\label{int_bur}}
We assume here that $\frac{\partial W_1}{\partial u_{n,m}}\neq0,\frac{\partial W_1}{\partial u_{n+k,m}}\neq0,\frac{\partial W_2}{\partial u_{n,m}}\neq0,\frac{\partial W_2}{\partial u_{n,m+l}}\neq0$ for at least  some $n,m$. The numbers $k,l$ are called the orders of these first integrals $W_1, W_2$, respectively: \seq{\ord W_1=k,\qquad \ord W_2=l.} 

It is clear due to (\ref{arb_fun}) that the orders of first integrals of a given equation are unbounded above. We are going to construct such examples for which the minimal (or lowest) orders of their first integrals may be arbitrarily high.

\section{Transition to the linear equation \eqref{lin_eq}}

In spite of the fact that the transform \eqref{hopf-cole} is not invertible, sometimes such transformations allow one to rewrite conservation laws and generalized symmetries from one equation to another, see e.g. \cite{y06} for the discrete-differential case. We are going to transfer first integrals from \eqref{ur_bur} to \eqref{lin_eq} and backward and to reduce in this way the problem to the case of the linear equation \eqref{lin_eq}.

Here we present four lemmas on the transfer first integrals from \eqref{ur_bur} to \eqref{lin_eq} and backward together with corresponding explicit formulae. More precisely, we discuss some relations between the first integrals (\ref{int_bur}) of eq. \eqref{ur_bur} and first integrals $\hat W_1, \hat W_2$ of eq. \eqref{lin_eq}: 
\eqs{\hat W_1=\hat w_{n,m}^{(1)}(v_{n,m},v_{n+1,m},\ldots,v_{n+\hat k,m}),\ \ \hat k\geq1,\\ \hat W_2=\hat w_{n,m}^{(2)}(v_{n,m},v_{n,m+1},\ldots,v_{n,m+\hat l}),\ \ \hat l\geq1.\label{int_lin}}

\begin{lemma}\label{lemW1} If eq. \eqref{ur_bur} possesses a first integral $W_1$ of an order $k$, then eq. \eqref{lin_eq} has a first integral $\hat W_1$ of the order $\hat k= k+1$.
\end{lemma}

\par\noindent {\bf Proof.}
Using the transform \eqref{hopf-cole}, we obtain
$$\hat W_1=w^{(1)}_{n,m}\left(\frac{v_{n+1,m}}{v_{n,m}},\frac{v_{n+2,m}}{v_{n+1,m}},\ldots,\frac{v_{n+k+1,m}}{v_{n+k,m}}\right).$$ It is clear that $\frac{\partial \hat W_1}{\partial v_{n,m}}\neq0,\frac{\partial \hat W_1}{\partial v_{n+k+1,m}}\neq0$ for at least some $n,m$, and therefore $\hat W_1$ not only is nontrivial (i.e. cannot depend on $n,m$ only) but also it has the order $k+1$. Due to \eqref{arb_fun} this order may become not minimal.

\begin{lemma}\label{lemW2} If eq. \eqref{ur_bur} possesses a first integral $W_2$ of an order $l$, then eq. \eqref{lin_eq} has a first integral $\hat W_2$ of an order $\hat l$, such that $1\leq \hat l\leq l$.
\end{lemma}

\par\noindent {\bf Proof.}
Using the transform \eqref{hopf-cole}, we are led to: 
$$\hat W_2=w^{(2)}_{n,m}\left(\frac{v_{n+1,m}}{v_{n,m}},\frac{v_{n+1,m+1}}{v_{n,m+1}},\ldots,\frac{v_{n+1,m+l}}{v_{n,m+l}}\right).$$ By induction on $l$ we can prove that 
$$v_{n+1,m+l}=\alpha^{nl}v_{n+1,m}-v_{n,m+l}+V_{n,l}(v_{n,m},v_{n,m+1},\ldots,v_{n,m+l-1}).$$ It is clear that $$\frac{\partial \hat W_2}{\partial v_{n,m+l}}=\frac{\partial w^{(2)}_{n,m}}{\partial u_{n,m+l}}\frac{\partial}{\partial v_{n,m+l}}\left(\frac{\alpha^{nl}v_{n+1,m}-v_{n,m+l}+\ldots}{v_{n,m+l}}\right)\neq 0$$ for some $n,m$, and therefore $\hat W_2$ is nontrivial. Its order is not greater than $l$.

\begin{lemma}\label{lemhatW1} If eq. \eqref{lin_eq} has a first integral $\hat W_1$ of an order $\hat k$, and $\hat W_1$ is linear w.r.t. $v_{n,m},v_{n+1,m},\ldots,v_{n+\hat k,m}$, then eq. \eqref{ur_bur} possesses a first integral $W_1$ of the order $k= \hat k$.
\end{lemma}

\par\noindent {\bf Proof.}
Using the transform \eqref{hopf-cole} and the property \eqref{arb_fun}, we obtain the relations
\eqs{\label{W1fromhat}W_1=\frac{T_1\hat W_1}{\hat W_1}=\frac{a_{0,n+1,m}v_{n+1,m}+a_{1,n+1,m}v_{n+2,m}+\ldots+a_{\hat k,n+1,m}v_{n+\hat k+1,m}}{a_{0,n,m}v_{n,m}+a_{1,n,m}v_{n+1,m}+\ldots+a_{\hat k,n,m}v_{n+\hat k,m}}\\=\frac{a_{0,n+1,m}v_{n+1,m}/v_{n,m}+a_{1,n+1,m}v_{n+2,m}/v_{n,m}+\ldots+a_{\hat k,n+1,m}v_{n+\hat k+1,m}/v_{n,m}}{a_{0,n,m}+a_{1,n,m}v_{n+1,m}/v_{n,m}+\ldots+a_{\hat k,n,m}v_{n+\hat k,m}/v_{n,m}}\\=\frac{a_{0,n+1,m}u_{n,m}+\ldots+a_{\hat k,n+1,m}u_{n+\hat k,m}u_{n+\hat k-1,m}\ldots u_{n,m}}{a_{0,n,m}+\ldots+a_{\hat k,n,m}u_{n+\hat k-1,m}u_{n+\hat k-2,m}\ldots u_{n,m}}.}
It is not difficult to check that $\frac{\partial W_1}{\partial u_{n,m}}\neq0,\frac{\partial  W_1}{\partial u_{n+\hat k,m}}\neq0$ for at least some values of $n,m$.

\begin{lemma}\label{lemhatW2} If eq. \eqref{lin_eq} has a first integral $\hat W_2$ of an order $\hat l$, and $\hat W_2$ is linear w.r.t. $v_{n,m},v_{n,m+1},\ldots,v_{n,m+\hat l}$, then eq. \eqref{ur_bur} possesses a first integral $W_2$ of the order $l= \hat l+1$.
\end{lemma}

\par\noindent {\bf Proof.}
By using the property \eqref{arb_fun} we obtain
\eqs{\label{W2fromhat}W_2=\frac{T_2\hat W_2}{\hat W_2}=\frac{b_{0,n,m+1}v_{n,m+1}+b_{1,n,m+1}v_{n,m+2}+\ldots+b_{\hat l,n,m+1}v_{n,m+1+\hat l}}{b_{0,n,m}v_{n,m}+b_{1,n,m}v_{n,m+1}+\ldots+b_{\hat l,n,m}v_{n,m+\hat l}}\\=\frac{b_{0,n,m+1}v_{n,m+1}/v_{n,m}+b_{1,n,m+1}v_{n,m+2}/v_{n,m}+\ldots+b_{\hat l,n,m+1}v_{n,m+1+\hat l}/v_{n,m}}{b_{0,n,m}+b_{1,n,m}v_{n,m+1}/v_{n,m}+\ldots+b_{\hat l,n,m}v_{n,m+\hat l}/v_{n,m}}\\=\frac{b_{0,n,m+1}Z_{n,m}+\ldots+b_{\hat l,n,m+1}Z_{n,m+\hat l}Z_{n,m+\hat l-1}\ldots Z_{n,m}}{b_{0,n,m}+\ldots+b_{\hat l,n,m}Z_{n,m+\hat l-1}Z_{n,m+\hat l-2}\ldots Z_{n,m}}.}
It follows from eqs. \eqref{hopf-cole} and \eqref{lin_eq} that
\seq{Z_{n,m}=\frac{v_{n,m+1}}{v_{n,m}}=\alpha^n\frac{u_{n,m}-1}{u_{n,m+1}+1}.}

\section{First integrals of the linear equation \eqref{lin_eq}}

We will use some necessary conditions of the Darboux integrability derived for the discrete case in \cite{as99}. Those conditions were formulated there for autonomous equations of the form \eqref{dis_eq}. We can reformulate and prove those conditions for the case of the non-autonomous linear equation \eqref{lin_eq}.

We can rewrite eq. \eqref{lin_eq} in the form
\eq{\label{ur_lin}(T_2-\alpha^n)(T_1+1)v_{n,m}+2\alpha^n v_{n,m}=0.} By using the discrete Laplace transformation \cite{as99}\eqs{\label{lap1}v_{n,m,j}=(T_1+\alpha^{j-1})v_{n,m,j-1},\ \ j\geq 1,\qquad v_{n,m,0}=v_{n,m},} we can introduce a sequence of unknown functions $v_{n,m,j}$  satisfying the equations
\eqs{\label{eq_vj1}(T_2-\alpha^{n+j})(T_1+\alpha^j)v_{n,m,j}+\alpha^{n+j}(1+\alpha^j) v_{n,m,j}=0 .} The last relations are proved by induction on $j$. One of the necessary conditions is formulated in terms of functions $K_{n,j}$:
\seqs{K_{n,j}=\alpha^{n+j}(1+\alpha^j),\ \ j\geq 1,\qquad K_{n,0}=2\alpha^n. }

The following theorem has been taken from \cite{as99}.

\begin{theorem} \label{hatW1} If eq. \eqref{ur_lin} possesses a first integral $\hat W_1$ of an order $\hat k $, then there exists $\tilde k$, $0\leq\tilde k< \hat k$, such that $K_{n,\tilde k}=0.$\end{theorem} 

In a similar way, we can rewrite eq. \eqref{lin_eq} in the form
\eq{\label{ur_lin2}(T_1+1)(T_2-\alpha^{n-1})v_{n,m}+\alpha^{n-1}(1+\alpha) v_{n,m}=0.} Using the second discrete Laplace transformation \eqs{\label{lap2}\check v_{n,m,j}=(T_2-\alpha^{n-j})\check v_{n,m,j-1},\ \ j\geq 1,\qquad \check  v_{n,m,0}=v_{n,m},} we can define a sequence of unknown functions $\check v_{n,m,j}$ which satisfy the equations
\eqs{\label{eq_vj2}(T_1+1)(T_2-\alpha^{n-j-1})\check v_{n,m,j}+\alpha^{n-j-1}(1+\alpha^{j+1})\check v_{n,m,j}=0 .} The last relations are proved by induction on $j$. The second of the necessary conditions is formulated in terms of functions $H_{n,j}$:
\seqs{H_{n,j}=\alpha^{n-j-1}(1+\alpha^{j+1}),\ \ j \geq 0. }

The following theorem has been taken from \cite{as99}.

\begin{theorem}\label{hatW2} If eq. \eqref{ur_lin2} possesses a first integral $\hat W_2$ of an order $\hat l$, then there exists $\tilde l$, $0\leq\tilde l< \hat l$, such that $H_{n,\tilde l}=0.$\end{theorem} 

Let $\alpha_N$ be a root of $-1$, such that:
\eq{\label{alph} \alpha_N^N=-1,\qquad \alpha_N^j\neq -1 ,\ \  1\leq j<N. } For any $N\geq1$, such a root always exists, for example, $\alpha_N=\exp(i\pi/N)$.
Let us now consider eq. \eqref{lin_eq} with $\alpha=\alpha_N$.
It follows from Theorems \ref{hatW1} and \ref{hatW2} for this equation that the orders of any its first integrals in the first and second directions must be such that:
\eq{\label{ord_hatW}\ord \hat W_1\geq N+1,\qquad \ord \hat W_2\geq N.} 

On the other hand, we can construct first integrals for eq. \eqref{lin_eq} with $\alpha=\alpha_N$ of such orders in an explicit form. Eqs. \eqref{eq_vj1} and \eqref{eq_vj2} with $j=N$ and $j=N-1$, respectively, take the form:
\eqs{(T_2+\alpha_N^n)(T_1-1)v_{n,m,N}=0,\\ (T_1+1)(T_2+\alpha_N^{n})\check v_{n,m,N-1}=0.\label{eq_vN}}
So we can find first integrals for these equations \eqref{eq_vN} in the first and second directions, respectively:
\eqs{\hat W_1=(-1)^m\alpha_N^{-nm}(T_1-1)v_{n,m,N},\\ \hat W_2=(-1)^n (T_2+\alpha_N^{n})\check v_{n,m,N-1}.}
By using the Laplace transformations \eqref{lap1} and \eqref{lap2}, we obtain first integrals for eq. \eqref{lin_eq} in the following explicit form:
\eqs{&\hat W_1=(-1)^m\alpha_N^{-nm}(T_1-1)(T_1+\alpha_N^{N-1})(T_1+\alpha_N^{N-2})\ldots(T_1+1)v_{n,m},\\
&\hat W_2=(-1)^n (T_2+\alpha_N^{n})(T_2-\alpha_N^{n-N+1})(T_2-\alpha_N^{n-N+2})\ldots(T_2-\alpha_N^{n-1})v_{n,m}.\label{hatW}}

We can see that both these first integrals are nontrivial, $\hat W_1$ is expressed in terms of  $v_{n,m},v_{n+1,m},\ldots, v_{n+N+1,m}$, while $\hat W_2$ is expressed in terms of $v_{n,m},v_{n,m+1},\ldots, v_{n,m+N}.$ Both $\hat W_1$ and $\hat W_2$ are linear w.r.t. $v_{n,m}$ and its shifts. We derive that \eqs{\label{ordalpha}\ord \hat W_1=N+1,\qquad\ord \hat W_2=N} and that these orders are minimal, taking into account the property \eqref{ord_hatW}.

For example, if $N=1$, then $\alpha_N= -1$, and we have first integrals of the minimal orders:
\seqs{\hat W_1=(-1)^{(1-n)m}(v_{n+2,m}-v_{n,m}),\qquad \hat W_2=(-1)^nv_{n,m+1}+v_{n,m}.} If $N=2$, then $\alpha_N=\pm i$, and the first integrals read:
\seqs{&\hat W_1=(-1)^m\alpha_N^{-nm}(v_{n+3,m}-v_{n+1,m}+\alpha_N(v_{n+2,m}-v_{n,m})),\\ &\hat W_2=(-1)^n(v_{n,m+2}+\alpha_N^{n-1}(\alpha_N-1)v_{n,m+1}-\alpha_N^{2n-1}v_{n,m}).} 
 
\section{First integrals of eq. \eqref{ur_bur}}

We consider here the equation \eqref{ur_bur} with $\alpha=\alpha_N$, where $\alpha_N$ is defined by (\ref{alph}). Using Lemmas \ref{lemhatW1} and \ref{lemhatW2} together with the formulae \eqref{W1fromhat} and \eqref{W2fromhat}, we construct first integrals $W_1$ and $W_2$ for eq. \eqref{ur_bur}, starting from the first integrals \eqref{hatW}. Their orders are: \eq{\ord  W_1=\ord  W_2=N+1.}
From Lemmas \ref{lemW1} and \ref{lemW2} and the relations \eqref{ordalpha} it follows that the minimal orders of first integrals of eq. \eqref{ur_bur} in both directions must be greater or equal than $N$. We are led to the following theorem:

\begin{theorem} Eq. \eqref{ur_bur} with $\alpha=\alpha_N$ is Darboux integrable. The minimal orders of its first integrals in both directions must be equal to $N$ or $N+1$.
\end{theorem}

It follows from this theorem that there exist Darboux integrable discrete equations among equations of the form \eqref{ur_bur}, such that the minimal orders of their first integrals in both directions are arbitrarily high.

For eq. \eqref{ur_bur} with $\alpha=\alpha_1=-1$, we obtain the following first integrals:
\seqs{W_1=(-1)^{-m}\frac{u_{n,m}(u_{n+2,m}u_{n+1,m}-1)}{u_{n+1,m}u_{n,m}-1},\quad  W_2=(-1)^n\frac{(u_{n,m+1}+u_{n,m+2})(u_{n,m}-1)}{(u_{n,m}+u_{n,m+1})(u_{n,m+2}+1)}.} In the case when $\alpha=\alpha_2=\pm i$, the first integrals read:
\seqs{ &W_1=(\alpha_2)^{-m}\frac{u_{n,m}(u_{n+3,m}u_{n+2,m}u_{n+1,m}+\alpha_2 u_{n+2,m}u_{n+1,m}-u_{n+1,m}-\alpha_2)}{u_{n+2,m}u_{n+1,m}u_{n,m}+\alpha_2 u_{n+1,m}u_{n,m}-u_{n,m}-\alpha_2},\\  &W_2=(\alpha_2)^n\frac{(u_{n,m}-1)(\alpha_2(u_{n,m+3}+u_{n,m+2})(u_{n,m+1}-1)-(u_{n,m+3}+1)(u_{n,m+2}+u_{n,m+1})}{(1+u_{n,m+3})(\alpha_2(u_{n,m+2}+u_{n,m+1})(u_{n,m}-1)-(u_{n,m+2}+1)(u_{n,m+1}+u_{n,m}))}.} 

In the paper \cite{gy12} a method is presented which uses so-called annihilation operators \cite{h05} and which allows one to construct first integrals of low orders and to show that those orders are minimal. By using this method we have checked that four first integrals given just above have the minimal orders. We believe that all first integrals of eq. \eqref{ur_bur} with $\alpha = \alpha_N$ which can be constructed by the scheme presented in this paper have the minimal orders.

As it has been said above, there is a hyperbolic equation \cite{ZS01} of the form \eqref{hyp}, namely,  \eq{\label{hyp_zs}u_{xy}=\frac{2N}{x+y}\sqrt{u_xu_y}} which is analogues to eq. \eqref{ur_bur} with $\alpha=\alpha_N$ in the sense that the minimal orders of first integrals of such equations \eqref{hyp_zs} may be arbitrarily high. Unlike eq. \eqref{hyp_zs}, which is symmetric under the involution $x\leftrightarrow y$, the discrete equation \eqref{ur_bur} is not symmetric under $n\leftrightarrow m$, and its first integrals in different directions have quite different forms. The second difference is that eq. \eqref{ur_bur} is of the Burgers type with linearizing transformation \eqref{hopf-cole}, while linearizing transformation for eq. \eqref{hyp_zs} has the form \seq{v=\sqrt{u_y},} where $v(x,y)$ is a solution of a hyperbolic linear equation.

\medskip
\subsection*{Acknowledgments.}
This work  has been supported by the Russian Foundation for Basic Research (grant
numbers: 10-01-00088-a, 11-01-97005-r-povolzhie-a, 12-01-31208-mol-a).

\end{document}